\documentclass[aps,prd,showkeys,nofootinbib,superscriptaddress]{revtex4-2}

\usepackage[autostyle,italian=guillemets]{csquotes}
\usepackage[utf8]{inputenc}
\usepackage[nowrite,infront,standard]{frontespizio}
\usepackage[english]{babel}
\usepackage{physics}
\usepackage{amsmath,amssymb}
\usepackage{xcolor}
\usepackage{mathrsfs}
\usepackage{subfloat}
\usepackage{frontespizio}
\usepackage{chemmacros}
\usepackage{float}
\usepackage{graphicx} 
\usepackage{bm}
\usepackage{appendix}
\usepackage{hyperref}
\usepackage{midpage}
\usepackage{braket}
\usepackage{placeins}
\usepackage{subfig}
\usepackage{slashed}

\def\ph{\varphi}

\def\al{a}\def\bt{b}

\def\te{\theta}
\def\si{\sigma}
\def\om{\omega}

\def\nh{_{(\!-\!)}}
\def\ph{_{(\!+\!)}}
\def\lc{{L}}
\def\rc{{R}}

\begin{document}
\title{Chiral and flavor oscillations in quantum field theory}


\author{Massimo Blasone}
\affiliation{Dipartimento di Fisica, Universit\`a degli Studi di Salerno, Via Giovanni Paolo II, 132 84084 Fisciano, Italy}
\affiliation{INFN, Sezione di Napoli, Gruppo Collegato di Salerno, Italy}

\author{Gennaro Zanfardino}
\affiliation{INFN, Sezione di Napoli, Gruppo Collegato di Salerno, Italy}
\affiliation{Dipartimento  di  Ingegneria Industriale, Universit\`a di Salerno, Via  Giovanni  Paolo  II, 132  I-84084  Fisciano  (SA),  Italy}
\affiliation{Institute of Nanotechnology of the National Research Council of Italy, CNR-NANOTEC, Rome Unit, Piazzale A. Moro 5, I-00185, Rome, Italy }

\date{\today}
%
\begin{abstract}
Massive fermions produced through weak charged interactions possess a definite chirality, which is not conserved during free propagation, leading to chiral oscillations. Neutrinos also exhibit flavor oscillations. These phenomena have been extensively studied—both individually and within a unified framework—using the first quantization approach, but only separately in the context of quantum field theory. In this work, we address this gap by developing a quantum field theory treatment of flavor-chiral oscillations. Focusing on the case of two-flavor mixing, we compute the expectation values of the (non-conserved) chiral-flavor charges and recover the same oscillation formulas obtained in the first quantization formalism. These charges are diagonalized via Bogoliubov transformations in both the chiral and flavor sectors. Such transformations reveal a non-trivial structure of the chiral-flavor vacuum, which consists of a condensate of particles with definite masses and helicities. 
\end{abstract}

\maketitle

\section{Introduction}
Fermions possess different degrees of freedom (DoFs), such as helicity and chirality \cite{WuTung,Pal,Pal:2010ih}. Helicity is defined as the projection of the spin onto the direction of the particle's momentum, and therefore depends on the particle's dynamics. In contrast, chirality is an intrinsic DoF associated with a specific symmetry of the complete Lorentz group. It is defined through the Dirac operator $\gamma^{5}$, which commutes with the Lorentz generators $\sigma^{\mu \nu}$. In other words, $\gamma^5$ enables the construction of projection operators onto irreducible subspaces of the Lorentz group, whose elements correspond to left- and right-chiral fields. 

While chirality is Lorentz invariant, helicity is not. For massive particles, a Lorentz boost in the direction opposite to their momentum can invert the momentum vector while leaving the spin unchanged, thereby flipping the helicity.

Moreover, although helicity is conserved during free evolution, chirality is not, since $\gamma^{5}$ does not commute with the mass term in the Dirac equation. As a consequence, a massive particle produced with definite chirality exhibits time-dependent chiral oscillations \cite{Fukugita:1994wx,Esposito:1998,Bernardini2005, Suekane:2021pdb, Bittencourt2021, Bittencourt2021_2, Bittencourt2022}, which can have observable effects in the non-relativistic regime\cite{Bittencourt2021_2,Ge2020}. This phenomenon arises in weak charged processes, where the interaction term involves a projection onto states with definite chirality \cite{Thomson:2013zua}.

Chirality and helicity only coincide for massless particles, for which helicity becomes  Lorentz invariant and chirality is conserved during time.

A particularly relevant case is that of neutrinos, which possess an additional intrinsic degree of freedom: flavor. Like chirality, flavor is not conserved over time. Flavor oscillations were first predicted in the pioneering works of Bruno Pontecorvo \cite{Pontecorvo}, and have since been extensively studied using a first quantization approach based on the relativistic Dirac equation \cite{Bernardini2004,Bernardini2005,Bernardini2006,Bernardini2011, Bittencourt2021, Bittencourt2023, Bittencourt2023_2,Nishi2006, Kimura:2021qlh}. These studies have explored flavor and chiral oscillations both separately and within a unified framework.  

In the non-relativistic regime, a general formula for the transition probability has been derived \cite{Bittencourt2021} (see Eq.\eqref{general_formula_oscillations} below), which includes three distinct contributions: the standard term originally obtained by Pontecorvo, a correction term arising from the Dirac bispinor structure, and a third term associated with the chiral component. Both correction terms become significant in the non-relativistic limit.

Similar results can also be obtained within the framework of second quantization. In particular, it is possible to define suitable non-conserved charges whose expectation values yield exactly the same oscillation formulas obtained in the first quantization approach. So far, these studies have been conducted separately for flavor oscillations \cite{bla95,bla98,Blasone:2001qa,Blasone:2002jv,Blasone:2003hh} and chiral oscillations for both Dirac and Majorana neutrinos \cite{Bittencourt2025, Morozumi:2025gmw,SalimAdam:2021suq}. The flavor and chiral charges are diagonalized through appropriate Bogoliubov transformations, revealing a non-trivial structure of the flavor and chiral vacua—both orthogonal to the standard Dirac vacuum. These results are also reviewed in \cite{Bittencourt2025_2}.

In this work, we unify the previously separate treatments of chiral and flavor oscillations into a single, coherent framework, thereby providing a more general theory of neutrino oscillations within QFT. In particular, we show that the generalized oscillation formula in Eq.\eqref{general_formula_oscillations} can be recovered by computing the expectation value of a chiral-flavor non-conserved charge. We also describe the associated Bogoliubov transformation and the resulting vacuum structure.

The paper is organized as follows: in Section, \ref{sec_bogoliubov_transformations}, we  review the Bogoliubov transformation that diagonalizes the chiral  charges. In Section \ref{sec_chiral_flavor_mixing} we extend this result by including  flavor mixing, and we analytically compute the expectation value of the chiral-flavor charge, obtaining the same formula for chiral and flavor oscillations as the one presented in Ref.\cite{Bittencourt2021} using a first quantization approach. We then analyze the structure of the   chiral-flavor vacuum in Section \ref{sec_chiral_flavor_vacuum}. Finally, in Section \ref{sec_conclusions}, we summarize our results and outline possible directions for future research.

\section{Charge diagonalization and Bogoliubov transformations} \label{sec_bogoliubov_transformations}
The Dirac Lagrangian
\begin{equation}
    \mathcal{L} \,= \,  \overline{\psi} \left( i \gamma^{\mu} \partial_{\mu} - m \right) \psi 
\end{equation}
is invariant under global phase transformations, which leads to the conserved charge
\begin{equation}
    Q = \int d^3{\bm x}\,\psi^{\dagger}(x)\psi(x).
\end{equation}
A Dirac bispinor field $\psi$ can be splitted into left and chiral parts as $\psi = \psi_{L} + \psi_{R}$ where
\begin{equation}
    \psi_{L} \equiv P_L \,\psi(x) \, =\, \frac{1 - \gamma^{5}}{2} \psi(x), \qquad  \psi_{R} \equiv P_R \,\psi(x) \, =\,  \frac{1 + \gamma^{5}}{2} \psi(x).
\end{equation}
and hence the Dirac Lagrangian can be written in terms of left and right chiral parts
\begin{equation}
    \mathcal{L}\, = \,\overline{\psi_{L}}   \,i\gamma^{\mu} \,\partial_{\mu}\, \psi_{L} \,+\,  \overline{\psi_{R}}  \,i\gamma^{\mu} \partial_{\mu} \psi_{R} - m\left(\overline{\psi_{L}}\,\psi_{R} \,+\, \overline{\psi_{R}}\,\psi_{L}\right).
\end{equation}
Because of the presence of the mass term, chiral symmetry is broken. Thus, separate global phase transformations for $\psi_{L}$ and $\psi_{R}$ lead to the non-conserved chiral charges 
\begin{equation}
    Q_{L}(t) = \int d^3{\bm x}\, \psi_{L}^\dagger (x)\psi_{L}(x), \qquad  Q_{R} (t)=  \int d^3{\bm x}\, \psi_{R}^\dagger(x) \psi_{R}(x).
\end{equation}
If the field is expressed in terms of the annihilation and creation operators $a_{\bm k, i}^{(s)}$ and  $a_{\bm k, i}^{(s) \dagger}$, which annihilate and create particles with momentum $\bm{k}$, helicity $s = \pm$, and mass $m$, we have
\begin{eqnarray}
\Psi(x) = \int d^{3}{\bm k} \, \sqrt{\frac{1}{2\omega_{k}}} 
\Bigg[
\Big( a_{\bm{k}}^{(+)}\, u_{+}(\bm{k}) + a_{\bm{k}}^{(-)}\, u_{-}(\bm{k}) \Big) e^{-i\omega_{k} t}
+ 
\Big( b_{-\bm{k}}^{(+) \dagger} v_{+}(-\bm{k}) + b_{-\bm{k}}^{(-) \dagger} v_{-}(-\bm{k}) \Big) e^{i\omega_{k} t}
\Bigg]
e^{i \bm{k} \cdot \bm{x}}
\end{eqnarray}
where the explicit form of the helicity basis spinors is given in Appendix \ref{app_chiral_basis_derivation}. 
The time-dependent chiral charges exhibit a non-diagonal form \cite{Bittencourt2025}, which can be diagonalized using Bogoliubov transformations \footnote{We determine the Bogoliubov coefficients by comparing the field expansion in different bases: with respect to  the analysis in \cite{Bittencourt2025}, which was restricted to a reference frame where the momentum $\bm{k}$ points along the $z$-axis, here we consider a generic direction for $\bm{k}$. }. 
In order to do so, we express the field in terms of the (orthonormal) chiral basis 
\begin{eqnarray}
    &&u_{L}(\bm k)= \frac{1}{\sqrt{2}} \left[\begin{matrix}-\sin{\left(\frac{\theta}{2} \right)}\\ 
e^{i \phi} \cos{\left(\frac{\theta}{2} \right)}\\ 
\sin{\left(\frac{\theta}{2} \right)}\\ 
- e^{i \phi} \cos{\left(\frac{\theta}{2} \right)}\end{matrix}\right], 
\,
u_{R}(\bm k) = \frac{1}{\sqrt{2}} \left[\begin{matrix}
\cos{\left(\frac{\theta}{2}\right)} \\
 e^{i \phi} \sin{\left(\frac{\theta}{2}\right)} \\
 \cos{\left(\frac{\theta}{2}\right)} \\
 e^{i \phi} \sin{\left(\frac{\theta}{2}\right)}
\end{matrix}\right], \notag \\
&&v_{L}(-\bm k) = \frac{1}{\sqrt{2}} \left[\begin{matrix}
 \cos{\left(\frac{\theta}{2}\right)} \\
 e^{i \phi} \sin{\left(\frac{\theta}{2}\right)} \\
-\cos{\left(\frac{\theta}{2}\right)} \\
- e^{i \phi} \sin{\left(\frac{\theta}{2}\right)}
\end{matrix}\right], \,
v_{R}(-\bm k) = \frac{1}{\sqrt{2}}\left[\begin{matrix} \sin{\left(\frac{\theta}{2} \right)}\\ -e^{i \phi} \cos{\left(\frac{\theta}{2} \right)}\\ 
\sin{\left(\frac{\theta}{2} \right)}\\ -e^{i \phi} \cos{\left(\frac{\theta}{2} \right)} \end{matrix}\right] , \label{chiral_bispinors}
\end{eqnarray}
which is discussed in Appendix \ref{app_chiral_basis_derivation}.

The Dirac field can be written as
\begin{eqnarray}
    \Psi(x) = \Psi_{L}(x) + \Psi_{R}(x)
\end{eqnarray}
with
\begin{eqnarray}\label{chiralexpansion}    \Psi_{L}(x) &=& \int d^{3}{\bm k}  \Big\{ a^{L}_{\bm{k}}(t) \, u_{L} + b^{L}_{-\bm{k}}(t) \, v_{L} \Big\} e^{-i \bm{k} \cdot \bm{x}} \, , \quad 
    \Psi_{R}(x) = \int d^{3}{\bm k}  \Big\{ a^{R}_{\bm{k}}(t) \, u_{R} + b^{R}_{-\bm{k}}(t) \, v_{R} \Big\} e^{-i \bm{k} \cdot \bm{x}}
\end{eqnarray}

where the transformed creation and annihilation operators are given by
\begin{eqnarray}
    a^{L}_{\bm k}(t) &\equiv& \cos{\sigma_{k}}\, e^{-i\omega_{k}t} \, a_{\bm k}^{\nh} \,+\, \sin{\sigma_{k}}\, e^{i\omega_{k}t}\, b_{-\bm k}^{\nh \dagger} \\[6pt]
    b^{L \dagger}_{-\bm k}(t) &\equiv& \cos{\sigma_{k}}\, e^{i\omega_{k}t}\, b_{-\bm k}^{\ph \dagger} \,+\, \sin{\sigma_{k}}\, e^{-i\omega_{k}t}\, a_{\bm k}^{\ph} \\[6pt]
    a^{R}_{\bm k}(t) &\equiv& \cos{\sigma_{k}}\, e^{-i\omega_{k}t}\, a_{\bm k}^{\ph} \,-\, \sin{\sigma_{k}}\, e^{i\omega_{k}t}\, b_{-\bm k}^{\ph \dagger} \\[6pt]
    b^{R \dagger}_{-\bm k}(t) &\equiv& \cos{\sigma_{k}}\, e^{i\omega_{k}t}\, b_{-\bm k}^{\nh \dagger} \,-\, \sin{\sigma_{k}}\, e^{-i\omega_{k}t}\, a_{\bm k}^{\nh}
\end{eqnarray}

and
\begin{eqnarray}
    \cos{\sigma_{k}} = \sqrt{\frac{1}{2}\left( 1 + \frac{k}{\omega_{k}}\right) } , \quad  \sin{\sigma_{k}} = \sqrt{ \frac{1}{2}\left( 1 - \frac{k}{\omega_{k}} \right)}\, .
\end{eqnarray}

The chiral charge is explicitly given by
\begin{eqnarray}
    Q_{L}(t) = \int d^{3}{\bm k} \Bigg\{ &&
    \cos^{2}{\sigma_{k}}\, a_{\bm k}^{\ph \dagger} a_{\bm k}^{\ph}
    + \sin^{2}{\sigma_{k}}\, a_{\bm k}^{\nh \dagger} a_{\bm k}^{\nh}
    - \cos^{2}{\sigma_{k}}\, b_{\bm k}^{\ph \dagger} b_{\bm k}^{\ph}
    - \sin^{2}{\sigma_{k}}\, b_{\bm k}^{\nh \dagger} b_{\bm k}^{\nh}
    \notag \\
    && +\, \sin(2\sigma_{k}) \Big[
    e^{2i\omega_{k}t} \big( a_{\bm k}^{\ph \dagger} b_{\bm k}^{\ph \dagger}
    + a_{\bm k}^{\nh \dagger} b_{\bm k}^{\nh \dagger} \big)
    + e^{-2i\omega_{k}t} \big( a_{\bm k}^{\ph} b_{\bm k}^{\ph}
    + a_{\bm k}^{\nh} b_{\bm k}^{\nh} \big)
    \Big] \Bigg\}
\end{eqnarray}

We find that the charge $Q_{L}(t)$ can be diagonalized in the form
\begin{eqnarray}
    Q_{L}(t) = \int d^{3}{\bm k} \Big( 
        a_{\bm k}^{L \dagger}(t)\, a_{\bm k}^{L}(t)
        - 
        b_{-\bm k}^{L \dagger}(t)\, b_{-\bm k}^{L}(t)
    \Big)
\end{eqnarray}
where the (left) chiral ladder operators are those appearing in the expansion of  $\Psi_{L}$ in the chiral basis, see Eq.\eqref{chiralexpansion}. Notice how the time dependence of the charge, is implicit in the chiral creation and annihilation operators. Similarly, for the right-chiral charge, we obtain
\begin{eqnarray}
    Q_{R}(t) = \int d^{3}{\bm k} \Big( 
        a_{\bm k}^{R \dagger}(t)\, a_{\bm k}^{R}(t)
        - 
        b_{-\bm k}^{R \dagger}(t)\, b_{-\bm k}^{R}(t)
    \Big)
\end{eqnarray}

These  results reproduce those of Ref.\cite{Bittencourt2025}, in an arbitrary reference frame. 

\section{Chiral-flavor oscillations} \label{sec_chiral_flavor_mixing}

We now consider flavor mixing:
\begin{eqnarray} \label{PontecorvoMix1}
\Psi_{e}(x) & = & \cos\theta\,\Psi_{1}(x) + \sin\theta\,\Psi_{2}(x)\\[2mm] \label{PontecorvoMix2}
\Psi_{\mu}(x) & = & -\sin\theta\,\Psi_{1}(x) + \cos\theta\,\Psi_{2}(x), \end{eqnarray}
where  $\Psi_{1}$ and $\Psi_{2}$ are fields  with masses $m_1$ and $m_2$.

Let us rewrite the chiral Bogoliubov transformations for different masses $m_{1}$ and $m_{2}$ as
\begin{eqnarray}
\label{transformation_chiral_flavor}
    a^{L}_{\bm k,i}(t) &\equiv& \cos{\sigma_{\bm k,i}}\, e^{-i\omega_{k,i}t}\, a_{\bm k,i}^{\nh}
    + \sin{\sigma_{\bm k,i}}\, e^{i\omega_{k,i}t}\, b_{-\bm k,i}^{\nh \dagger} \\[6pt]
    b^{L \dagger}_{-\bm k,i}(t) &\equiv& \cos{\sigma_{\bm k,i}}\, e^{i\omega_{k,i}t}\, b_{-\bm k,i}^{\ph \dagger}
    + \sin{\sigma_{\bm k,i}}\, e^{-i\omega_{k,i}t}\, a_{\bm k,i}^{\ph} \\[6pt]
    a^{R}_{\bm k,i}(t) &\equiv& \cos{\sigma_{\bm k,i}}\, e^{-i\omega_{k,i}t}\, a_{\bm k,i}^{\ph}
    - \sin{\sigma_{\bm k,i}}\, e^{i\omega_{k,i}t}\, b_{-\bm k,i}^{\ph \dagger} \\[6pt]
    b^{R \dagger}_{-\bm k,i}(t) &\equiv& \cos{\sigma_{\bm k,i}}\, e^{i\omega_{k,i}t}\, b_{-\bm k,i}^{\nh \dagger}
    - \sin{\sigma_{\bm k,i}}\, e^{-i\omega_{k,i}t}\, a_{\bm k,i}^{\nh}
\end{eqnarray}
where
\begin{eqnarray}
    \cos^{2}{\sigma_{{\bm k},i}} &=& \frac{1}{2}\left( 1 + \frac{|{\bm k}|}{\omega_{k,i}}\right), \quad \sin^{2}{\sigma_{{\bm k},i}} = \frac{1}{2}\left( 1 - \frac{|{\bm k}|}{\omega_{k,i}}\right), \quad \omega_{k,i} =\sqrt{|{\bm k}|^{2}+m_{i}^{2}}
\end{eqnarray}
for $i=1,2$.

We introduce the  left and right flavor/chiral operators as
\begin{eqnarray}\label{chiflav_transf1}
\al_{{\bm k},e}^{I} (t)  &= \, \!\cos\te\,\al_{{\bm k},1}^{I}(t) + \sin\te
\,\al^{I}_{{\bm k}, 2}(t), \quad
&\bt_{-{\bm k},e}^{I} (t)  = \,\cos\te\,\bt_{-{\bm k},1}^{I}(t) + \sin\te
\, \bt^{I}_{-{\bm k}, 2}(t) ,   \\ 
\al_{{\bm k},\mu}^{I} (t)  &= \, \!-\sin\te\,\al_{{\bm k},1}^{I}(t) + \cos\te
\, \al^{I}_{{\bm k}, 2}(t), \quad
&\bt_{-{\bm k},\mu}^{I} (t)  = \,-\sin\te\,\bt_{-{\bm k},1}^{I}(t) + \cos\te
\, \bt^{I}_{-{\bm k}, 2}(t)
\label{chiflav_transf2}
\end{eqnarray}
for $I=L,R$. 

The CAR for the transformed operators hold:
\begin{equation}
  \{ a_{{\bm k},\si}^{I} (t) ,a_{{\bm k},\rho}^{J \dagger} (t)\} \,= \,\delta_{\si \rho} \delta_{I J} \, ,\quad
     \{ a_{{\bm k},\si}^{I} (t) ,a_{{\bm k},\rho}^{J} (t)\} \,= \,\{ \al_{{\bm k},\si}^{I} (t) ,\bt_{-{\bm k},\rho}^{J} (t)\} \,=\, \{ b_{{\bm k},\si}^{I} (t) ,b_{{\bm k},\rho}^{J} (t)\} \,=\, 0
\end{equation}
and similar ones with $\si,\rho= e,\mu$ and $I,J=R,L$.

We can check that the flavor fields can be written as
\begin{eqnarray}\label{flavchirexp_e}    \Psi_e(x) & =&  \int d^{3}{\bm k}  \Big\{ a^{L}_{\bm{k},e}(t) \, u_{L}(\bm k) + b^{L}_{-\bm{k},e}(t) \, v_{L}(-\bm k)  \, +\,a^{R}_{\bm{k},e}(t) \, u_{R}(\bm k) + b^{R}_{-\bm{k},e}(t) \, v_{R}(-\bm k) \Big\} e^{-i \bm{k} \cdot \bm{x}}\,,
\\ [2mm] \label{flavchirexp_mu}
 \Psi_\mu(x) & =&  \int d^{3}{\bm k}  \Big\{ a^{L}_{\bm{k},\mu}(t) \, u_{L}(\bm k) + b^{L}_{-\bm{k},\mu}(t) \, v_{L}(-\bm k)  \, +\,  a^{R}_{\bm{k},\mu}(t) \, u_{R}(\bm k) + b^{R}_{-\bm{k},\mu}(t) \, v_{R}(-\bm k) \Big\} e^{-i \bm{k} \cdot \bm{x}}\,.
\end{eqnarray}

\subsection*{Chiral-flavor charge expectation value and survival probability} \label{sec_chiral_flavor_exp_value}

By means of these operators, operating in a similar way as done for the case of separate chiral and flavor oscillations in Refs.\cite{bla98,Bittencourt2025}, the complete formula for chiral-flavor neutrino oscillations can be derived. 

The chiral-flavor charge, in its diagonal form is given by
\begin{eqnarray}
    Q^{L}_{e}(t) = \int d^{3} {\bm k } \left( {\al_{{\bm k},e}^{L ^{\dagger}}}(t) \al_{{\bm k},e}^{L}(t)  - {\bt_{-{\bm k},e}^{\lc ^{\dagger} }}(t) \bt_{-{\bm k},e}^{\lc} (t)\right) 
\end{eqnarray}
and its expectation value on a  left-chiral electron neutrino state can be computed as
\begin{eqnarray}
    \braket{Q_{e}^{L}(t)} = \braket{a_{\bm k,e}^{L}|Q_{e}^{L}(t)|a_{\bm k,e}^{L}}
\end{eqnarray}
where
\begin{eqnarray}
    \ket{\al^{\lc}_{{\bm k},e}} \equiv a_{\bm k, e}^{L \dagger}\ket{\tilde{0}},
\end{eqnarray}
$a_{\bm k, e}^{L } \equiv a_{\bm k, e}^{L}(0)$ is taken at time $t=0$, and $\ket{\tilde{0}}$ is the chiral-flavor vacuum which is annihilated by  $a_{\bm k, e}^{L}$.
\begin{eqnarray}
     \braket{Q_{e}^{L}(t)} = \braket{\tilde{0}|a_{\bm k,e}^{L} a_{\bm k,e}^{L\dagger}(t) a_{\bm k,e}^{L}(t) a_{\bm k,e}^{L\dagger}|\tilde{0}}
     -   \braket{\tilde{0}|a_{\bm k,e}^{L} b_{\bm k,e}^{L\dagger}(t) b_{\bm k,e}^{L}(t) a_{\bm k,e}^{L\dagger}|\tilde{0}}
\end{eqnarray}

Note that $\{a_{\bm k,e}^{L}, a_{\bm k,e}^{L}(t) \} = 0$, $\{a_{\bm k,e}^{L}, b_{\bm k,e}^{L}(t) \} = 0$ and $\{a_{\bm k,e}^{L}, b_{\bm k,e}^{L \dagger}(t) \} = 0$. Therefore, by using the CAR at equal time, we obtain
\begin{eqnarray}
    \braket{Q_{e}^{L}(t)} = \left| \left\{ a_{\bm k,e}^{L}(0), a_{\bm k,e}^{L \dagger}(t)\right\} \right|^{2} 
    +  \braket{\tilde{0}|a_{\bm k,e}^{L \dagger}(t) a_{\bm k,e}^{L}(t)|\tilde{0}}
    -  \braket{\tilde{0}|b_{\bm k,e}^{L \dagger}(t) b_{\bm k,e}^{L}(t)|\tilde{0}}
    \label{eq_chiral_expectation}
\end{eqnarray}
BY means of the transformations Eq.~\eqref{transformation_chiral_flavor} and their inverses at $t=0$, one can check that the last two terms in Eq.~\eqref{eq_chiral_expectation} cancel out, and only the anticommutator term survives. We obtain
\begin{eqnarray}
    \left\{ a_{\bm k,e}^{L}(0), a_{\bm k,e}^{L \dagger}(t)\right\} &=& \cos^{2}{\theta} \left( e^{i\omega_{1}t}\cos^{2}{\sigma}_{\bm k,1}+ e^{-i \omega_{1}t}\sin^{2}{\sigma}_{\bm k,1 }\right) + \sin^{2}{\theta} \bigg( e^{i\omega_{2}t}\cos^{2}{\sigma}_{\bm k,2} + e^{-i\omega_{2}t}\sin^{2}{\sigma}_{\bm k,2}  \bigg)  
    \label{anticomm_flavor_chiral}
\end{eqnarray}

which leads to (see Appendix~\ref{app_proof_prob_electronic}).

\begin{equation}
     \braket{Q_{e}^{L}(t)} \equiv P_{e \to e}(t) = P_S^{e \to e}(t) + A_e(t) + B_e(t).
\end{equation}
Where $ P_S^{e \to e}(t) $ is the standard flavor oscillation formula:
\begin{equation}
    P_S^{e \to e}(t) = 1 - \sin^2(2\theta) \sin^2 \left( \frac{\omega_{2}- \omega_{1}}{2} t \right),
\label{general_formula_oscillations}
\end{equation}
and
\begin{equation}
    A_e(t) = - \left( \frac{m_1}{\omega_{k,1}} \cos^2(\theta) \sin(\omega_{k,1} t) + \frac{m_2}{\omega_{k,2}} \sin^2(\theta) \sin(\omega_{k,2} t) \right)^2,
\end{equation}

\begin{equation}
    B_e(t) = \frac{1}{2} \sin^2(2\theta) \sin(\omega_{k,1} t) \sin(\omega_{k,2} t) 
    \left( \frac{k^2 + m_1 m_2}{\omega_{k,1} \omega_{k,2}} - 1 \right),
\end{equation}
which exactly matches the formula obtained in Ref.\cite{Bittencourt2021_2}: this is the main result of the present work.

\section{Chiral-flavor vacuum} \label{sec_chiral_flavor_vacuum}

We now consider the structure of the vacuum state for the chiral/flavor annihilation operators defined in Eqs.\eqref{chiflav_transf1},\eqref{chiflav_transf2}. By following Ref.\cite{Bittencourt2025}, we write:
\begin{eqnarray}\label{flavmass_mass1}
   \al_{{\bm k},i}^{\lc} (t)&=
    \widetilde{G}^{-1}_t \al_{{\bm k},i}^{\nh} (t)\widetilde{G}_t \qquad, \quad \bt_{-{\bm k, i}}^{\lc}(t) \,&=\,
    \widetilde{G}^{-1}_t \bt^{\ph}_{-{\bm k},i} (t)\widetilde{G}_t
    \\ [2mm]
    \label{flavmass_mass2}
     \al_{{\bm k},i}^{\rc}(t) &=
    \widetilde{G}^{-1}_t \, \al_{{\bm k},i}^{\ph} (t)\,\widetilde{G}_t\qquad, \quad \bt_{-{\bm k},i}^{\rc}(t) \,&=\,
    \widetilde{G}^{-1}_t \,\bt_{-{\bm k, i}}^{\nh}(t) \,\widetilde{G}_t,
\end{eqnarray}
with $i=1,2$ and   the generator $ \widetilde{G}_t$ given by
\begin{equation}
\label{generator_transformations}
     \widetilde{G}_t(\theta,\phi)\, = \, 
    \exp \left[\sum_i\sum_r\int d^3{\bm k} \, \sigma_k \epsilon^r \left( e^{- 2 i \om_{k,i} t}a_{{\bm k},i}^{(r)} b_{-{\bm k},i}^{(r)} - 
    e^{2 i \om_{k,i} t}b_{-{\bm k},i}^{(r) \dagger} a_{{\bm k},i}^{(r) \dagger} \right)\right].
\end{equation}
The action of the generator $ \widetilde{G}_t(\theta,\phi)$  on the vacuum $|0\rangle$ defines the \emph{massive chiral vacuum} \cite{Bittencourt2025}:
\begin{equation}
    |{\tilde 0} (t) \rangle \, \equiv \widetilde{G}^{-1}_t 
 \,  |0\rangle  \, = \, \prod_i\prod_{{\bm k},r}\Big[\cos{\sigma_{k,i}} \,+\, \epsilon^r e^{ i2 i \om_{k,i} t} \sin{\sigma_{k,i}} \, a_{{\bm k},i}^{(r) \dagger}b_{-{\bm k}, i}^{(r) \dagger}\Big]| 0\rangle.
\end{equation}

Flavor-chiral operators are written in a compact way as
\begin{eqnarray}\label{flavchi_mass1}
   \al_{{\bm k},\sigma}^{L} (t)&=
   \widetilde{G}^{-1}_t R^{-1}_\theta \, \al_{{\bm k},i}^{\nh} (t)\, R_\theta \,\widetilde{G}_t \qquad, \quad &\bt_{{\bm k,\sigma}}^{L}(t) \,=\,
    \widetilde{G}^{-1}_t R^{-1}_\theta \,\bt^{\ph}_{{\bm k},i} (t) \,R_\theta\,\widetilde{G}_t
    \\ \label{flavchi_mass2}
    \al_{{\bm k},\sigma}^{R} (t)&=
   \widetilde{G}^{-1}_t R^{-1}_\theta \, \al_{{\bm k},i}^{\ph} (t)\, R_\theta \,\widetilde{G}_t \qquad, \quad &\bt_{{\bm k,\sigma}}^{R}(t) \,=\,
    \widetilde{G}^{-1}_t R^{-1}_\theta \,\bt^{\nh}_{{\bm k},i} (t) \,R_\theta\,\widetilde{G}_t
\end{eqnarray}
with $(\sigma, i)=(e,1), (\mu,2)$. Here $R_\theta$ is the generator of rotations among annihilation operators with different masses:
\begin{equation}
    R_\theta\equiv \exp[   \theta  \sum_r\int d^3{\bm k} \,  \Big( a_{{\bm k},1}^{(r)\dagger}a_{{\bm k},2}^{(r)} \, +\, b_{{\bm k},1}^{(r)\dagger}b_{{\bm k},2}^{(r)} \,- \, a_{{\bm k},2}^{(r)\dagger}a_{{\bm k},1}^{(r)}  \,- \, b_{{\bm k},2}^{(r)\dagger}b_{{\bm k},1}^{(r)} \Big) ] 
\end{equation}
Since the action of the above rotation on the vacuum $|0\rangle$ is trivial, we obtain that the vacuum state for the flavor-chiral operators coincides with the massive chiral vacuum:
\begin{equation}
    \al_{{\bm k},\sigma}^{I}(t)  |{\tilde 0} (t) \rangle  \, = \, \widetilde{G}^{-1}_t R^{-1}_\theta \, \al_{{\bm k},i}^{\nh} (t)\, R_\theta \,\widetilde{G}_t |{\tilde 0} (t) \, = \, \widetilde{G}^{-1}_t R^{-1}_\theta \, \al_{{\bm k},i}^{\nh} (t)\, R_\theta \, |0\rangle \, = \,0
\end{equation}
because $R_\theta \, |0\rangle = |0\rangle$.

By means of the inverses of Eqs.\eqref{flavmass_mass1},
\eqref{flavmass_mass2}, Eqs.\eqref{flavchi_mass1},\eqref{flavchi_mass2} can be written as
\begin{eqnarray}\label{flavchi_flav}
   \al_{{\bm k},\sigma}^{I} (t)&=
   \widetilde{R}^{-1}_\theta (t) \, \al_{{\bm k},\si}^{I} (t)\, \,\widetilde{R}_\theta(t) \qquad, \quad &\bt_{{\bm k,\sigma}}^{I}(t) \,=\,
   \widetilde{R}^{-1}_\theta (t)\,\bt^{I}_{{\bm k},\si} (t) \,\widetilde{R}_\theta(t)
    \end{eqnarray}
in agreement with Eqs.\eqref{chiflav_transf1}, \eqref{chiflav_transf2}.

Here we have defined the transformed rotation generator $ \widetilde{R}_\theta(t) \equiv \widetilde{G}^{-1}_t R_\theta \,\widetilde{G}_t$, acting on the chiral ladder operators (at time $t$):
\begin{equation}
   \widetilde{R}_\theta(t)\, =\, \exp[   \theta  \sum_I\int d^3{\bm k} \,  \Big( a_{{\bm k},1}^{I\dagger}(t)a_{{\bm k},2}^{I}(t) \, +\, b_{{\bm k},1}^{I\dagger}(t)b_{{\bm k},2}^{I}(t) \,- \, a_{{\bm k},2}^{I\dagger}(t)a_{{\bm k},1}^{I}(t)  \,- \, b_{{\bm k},2}^{I \dagger}(t)b_{{\bm k},1}^{I}(t) \Big) ] 
\end{equation}
Of course, $ \widetilde{R}_\theta(t)$ does not leave the vacuum $|0\rangle$ invariant.

\section{Conclusions} \label{sec_conclusions}
In this work, we have developed a unified quantum field theoretical framework for describing chiral and flavor oscillations of fermions, with a particular focus on neutrinos. Building upon previous studies that treated chiral and flavor oscillations separately, we introduced  non-conserved chiral-flavor charges whose expectation value encapsulates the dynamics of both phenomena simultaneously.

We demonstrated that the chiral-flavor charges can be diagonalized via an appropriate Bogoliubov transformation, resulting in  a non-trivial vacuum structure, with the massive chiral vacuum being orthogonal to the standard Dirac vacuum. By computing the expectation value of this charge, we recovered the general oscillation formula previously derived in Ref.\cite{Bittencourt2021} within a first quantization formalism. This result provides a  consistency check for our approach and reinforces the equivalence between the first and second quantization treatments in the appropriate regimes.

Our findings highlight the significance of vacuum structure in neutrino oscillations and offer a deeper understanding of the interplay between different fermionic degrees of freedom, especially in the non-relativistic regime where chiral corrections become relevant.

Future work could explore the extension of this formalism to three-flavor neutrino mixing, incorporate CP-violating phases, and analyze the implications of external fields or matter effects \cite{Li:2023iys} on the unified chiral-flavor dynamics. Another interesting question is how to reproduce the non-perturbative results of this paper in the framework of perturbative finite-time approach of Refs.\cite{Blasone:2023brf,Blasone:2024zsn,Blasone:2024nuf,ChiralInt}.

\section*{Acknowledgements}

The authors thank V. A. S. Bittencourt (University of Strasbourg) for insightful discussions.

\appendix 

\section*{Appendix}

\subsection{Chiral bispinors derivation} \label{app_chiral_basis_derivation}
We start from the helicity basis \cite{Thomson:2013zua}
\begin{eqnarray}
&u_{+}(\bm k)=\sqrt{\omega_{k}+m}\;
\left(\begin{array}{c} \cos{\theta}  \\ \sin{\theta} \,e^{i\phi} \\ \frac{k}{\omega_{k}+m} \,\cos{\theta}  \\ \frac{k}{\omega_{k}+m} \,\sin{\theta} \,e^{i\phi}  \end{array}\right) \, , \quad
&u_{-}(\bm k)=\sqrt{\omega_{k}+m}\;
\left(\begin{array}{c} - \sin{\theta}  \\ \cos{\theta} \,e^{i\phi} \\ \frac{k}{\omega_{k}+m} \,\sin{\theta}  \\ -\frac{k}{\omega_{k}+m} \,\cos{\theta}\, e^{i\phi}  \end{array}\right) \\ [2mm]
&v_{+}(-\bm k)=\sqrt{\omega_{k}+m}\;
\left(\begin{array}{c} \frac{k}{\omega_{k}+m} \,\cos{\theta}  \\  \frac{k}{\omega_{k}+m} \,\sin{\theta} \,e^{i\phi}
\\
-\cos{\theta}  \\-\sin{\theta} \,e^{i\phi}  \end{array}\right) \, , \quad
&v_{-} (-\bm k)=\sqrt{\omega_{k}+m}\;
\left(\begin{array}{c} \frac{k}{\omega_{k}+m} \,\sin{\theta}  \\  -\frac{k}{\omega_{k}+m} \,\cos{\theta} \,e^{i\phi}
\\
\sin{\theta}  \\- \cos{\theta} \,e^{i\phi}   \end{array}\right)
\end{eqnarray}
where the momentum is expressed in spherical coordinates $\bm k = (k \sin{\theta} \cos{\phi}, k\sin{\theta}\sin{\phi}, k \cos{\theta})$. The basis is orthogonal with 
\begin{eqnarray}
u_{r}^{\dagger}(\bm k)u_{s}(\bm k) = 2\omega_{k} \delta_{rs}\, , \quad v_{r}^{\dagger}(\bm k)v_{s}(\bm k)= 2\omega_{k} \delta_{rs}\, , \quad u_{r}^{\dagger}(\bm k)v_{s}(-\bm k)=0\, , \qquad \text{for } r,s= +,-
 \end{eqnarray}



We define the orthonormal chiral bispinors as
\begin{eqnarray}
    &u_{L}(\bm k) = \frac{1}{\sqrt{2}} \left[\begin{matrix}-\sin{\left(\frac{\theta}{2} \right)}\\ 
e^{i \phi} \cos{\left(\frac{\theta}{2} \right)}\\ 
\sin{\left(\frac{\theta}{2} \right)}\\ 
- e^{i \phi} \cos{\left(\frac{\theta}{2} \right)}\end{matrix}\right]\,, 
\quad
&u_{R}(\bm k) = \frac{1}{\sqrt{2}} \left[\begin{matrix}
\cos{\left(\frac{\theta}{2}\right)} \\
 e^{i \phi} \sin{\left(\frac{\theta}{2}\right)} \\
 \cos{\left(\frac{\theta}{2}\right)} \\
 e^{i \phi} \sin{\left(\frac{\theta}{2}\right)}
\end{matrix}\right], \\ [2mm]
&v_{L}(-\bm k) = \frac{1}{\sqrt{2}} \left[\begin{matrix}
 \cos{\left(\frac{\theta}{2}\right)} \\
 e^{i \phi} \sin{\left(\frac{\theta}{2}\right)} \\
-\cos{\left(\frac{\theta}{2}\right)} \\
- e^{i \phi} \sin{\left(\frac{\theta}{2}\right)}
\end{matrix}\right]\,, 
\quad
& v_{R}(-\bm k) = \frac{1}{\sqrt{2}}\left[\begin{matrix} \sin{\left(\frac{\theta}{2} \right)}\\ -e^{i \phi} \cos{\left(\frac{\theta}{2} \right)}\\ 
\sin{\left(\frac{\theta}{2} \right)}\\ -e^{i \phi} \cos{\left(\frac{\theta}{2} \right)} \end{matrix}\right] 
\label{chiral_flavor_basis}
\end{eqnarray}
the same as in Eq. \ref{chiral_bispinors}.
Clearly, since $P_{L(R)}$ is a projector, $u_{L(R)}$ and $v_{L(R)}$ are its eigenstates. Therefore, we can write 
\begin{eqnarray}
    &u_{+}(\bm k)= \sqrt{2\omega_{k}} \left( \cos{\theta_{p}u_{R}} + \sin{\theta_{p} v_{L}} \right)\, , \quad 
    &u_{-}(\bm k) = \sqrt{2\omega_{k}}\left( \cos{\theta_{k}} u_{L} - \sin{\theta_{k}}v_{R} \right)  \\[2mm]
    &v_{-}(-\bm k)= \sqrt{2\omega_{k}} \left( \cos{\theta_{k}v_{L}} - \sin{\theta_{k} u_{R}} \right)\, , \quad 
    &v_{-}(-\bm k) = \sqrt{2\omega_{k}}\left( \cos{\theta_{k}} v_{R} - \sin{\theta_{k}}u_{L} \right) 
\end{eqnarray}
and inverses
\begin{eqnarray}
    &u_{L}(\bm k) = \frac{1}{\sqrt{2\omega_{k}}}\left( \cos{\theta_{k}}u_{-}(\bm k) + \sin{\theta_{k}} v_{-}(-\bm k) \right)\,, \quad &u_{R}(\bm k) = \frac{1}{\sqrt{2\omega_{k}}}\left( \cos{\theta_{k}}u_{+}(\bm k) - \sin{\theta_{k}} v_{+}(-\bm k) \right) \\
    &v_{L}(-\bm k) = \frac{1}{\sqrt{2\omega_{k}}}\left( \cos{\theta_{k}}v_{+}(-\bm k) + \sin{\theta_{k}} u_{+}(\bm k) \right)\,, \quad &v_{R}(- \bm k) = \frac{1}{\sqrt{2\omega_{k}}}\left( \cos{\theta_{k}}v_{-}(-\bm k) - \sin{\theta_{k}} u_{-}(\bm k) \right)
\end{eqnarray}
 Note also that for $m=0$, it is $\sin{\theta_{k}}=0$ and $\cos{\theta_{k}}=1$. Thus the above relations are consistent with the requirement tha for massless particles   helicity coincides with chirality. The orthonormality relations for the  basis in Eq. \eqref{chiral_flavor_basis} are:
 \begin{eqnarray}
     u_{I}^{\dagger}(\bm k)u_{J}(\bm k) = \delta_{IJ}\, , \quad v_{I}^{\dagger}(\bm k)v_{J}(\bm k)=\delta_{IJ}\, , \quad u_{I}^{\dagger}(\bm k)v_{J}(-\bm k)=0\, , \qquad \text{for } I,J=L,R
 \end{eqnarray}

\subsection{Explicit derivation of the survival probability of an electronic and left-chiral neutrino} \label{app_proof_prob_electronic}

Starting from the anti-commutator \eqref{anticomm_flavor_chiral}, since $(a+b)(a^{*}+b^{*}) = |a|^{2}+|b|^{2} + 2 \text{Re}(ab^{*})$ for any complex numbers $a$ and $b$, we can write 
\begin{eqnarray}
     \left|\left\{ a_{\bm k,e}^{L}(0), a_{\bm k,e}^{L \dagger}(t)\right\}\right|^{2} &=& \cos^{4}\theta \left[\cos^{4}\sigma_{\bm k, 1} + \sin^{4}\sigma_{\bm k, 1} + 2 \text{Re} \left( \cos^{2}\sigma_{\bm k, 1}\sin^{2}\sigma_{\bm k, 1} e^{2i\omega_{k,1}t} \right)\right] \notag \\
     &&+\sin^{4}\theta \left[\cos^{4}\sigma_{\bm k, 2} + \sin^{4}\sigma_{\bm k, 2} + 2 \text{Re} \left( \cos^{2}\sigma_{\bm k, 2}\sin^{2}\sigma_{\bm k, 2} e^{-2i\omega_{k,2}t} \right) \right]\notag \\
     && +  \frac{\sin^{2}{2\theta}}{2} \text{Re} \bigg\{ \left( e^{-i\omega_{k,1}t}\cos^{2}\sigma_{\bm k, 1} + e^{i\omega_{k,1}t} \sin^{2}\sigma_{\bm k, 1}\right) \left( e^{i\omega_{l,2}t}\cos^{2}\sigma_{\bm k, 2} +   \sin^{2}\sigma_{\bm k, 2} e^{-i\omega_{k,2} t} \right) \bigg\} \label{anticomm_squared_2}
\end{eqnarray}
Since
\begin{eqnarray}
    2 \text{Re} \left( \cos^{2}\sigma_{\bm k, i}\sin^{2}\sigma_{\bm k, i} e^{-i2\omega_{i}t}\right) = \frac{m_{i}}{2 \omega_{i}^{2}} \cos{(2 \omega_{k,i} t)} = \frac{m_{i}^{2}}{2 \omega_{k,i}^{2}} - \frac{m_{i}^{2}}{ \omega_{k,i}^{2}} \sin^{2}{(\omega_{k,i}t)}
\end{eqnarray}
and
\begin{eqnarray}
    \cos^{4}\sigma_{\bm k, i} + \sin^{4}\sigma_{\bm k, i}  = \frac{1}{2} + \frac{k^{2}}{2 \omega_{k,i}^{2}},
\end{eqnarray}
Eq. \eqref{anticomm_squared_2} can be rewritten as
\begin{eqnarray}
    \left|\left\{ a_{\bm k,e}^{L}(0), a_{\bm k,e}^{L \dagger}(t)\right\}\right|^{2} &=& 
    \cos^{4}\theta \left[1 - \frac{m_{1}^{2}}{\omega_{k,1}^{2}}\sin^{2}{(\omega_{k,1}t)}\right]+\sin^{4}\theta \left[1 - \frac{m_{2}^{2}}{\omega_{k,2}^{2}}\sin^{2}{(\omega_{k,2}t)}\right] \notag \\
     && +  \frac{\sin^{2}{2\theta}}{2} \text{Re} \bigg\{ \left( e^{-i\omega_{k,1}t}\cos^{2}\sigma_{\bm k, 1} + e^{i\omega_{k,1}t} \sin^{2}\sigma_{\bm k, 1}\right) \left( e^{i\omega_{k,2}t}\cos^{2}\sigma_{\bm k, 2} +   \sin^{2}\sigma_{\bm k, 2} e^{-i\omega_{k,2} t} \right) \bigg\}
\end{eqnarray}

The real part of the last term in the brackets can be rewritten as
\begin{eqnarray}
&& \sin^{2}\sigma_{\bm{k},2} \cos^{2}\sigma_{\bm{k},1} \cos{[(\omega_{k,1}+\omega_{k,2})t]} + \cos^{2}\sigma_{\bm{k},2} \cos^{2}\sigma_{\bm{k},1} \cos{[(\omega_{k,2}-\omega_{k,1})t]} \notag \\
&& + \sin^{2}\sigma_{\bm{k},2} \sin^{2}\sigma_{\bm{k},1} \cos{[(\omega_{k,2}-\omega_{k,1})t]} + \cos^{2}\sigma_{\bm{k},2} \sin^{2}\sigma_{\bm{k},1} \cos{[(\omega_{k,1}+\omega_{k,2})t]} \notag \\
&& = \cos{\omega_{k,1}t} \cos{\omega_{k,2}t} + \sin{\omega_{k,1}t} \sin{\omega_{k,2}t} \left[ (\cos^{2}\sigma_{\bm{k},1} - \sin^{2}\sigma_{\bm{k},1})(\cos^{2}\sigma_{\bm{k},2} - \sin^{2}\sigma_{\bm{k},2}) \right] \notag
\end{eqnarray}
Therefore,
\begin{eqnarray}
    \braket{Q_{e}^{L}(t)} &=& 
    \cos^{4}\theta \left[1 - \frac{m_{1}^{2}}{\omega_{k,1}^{2}}\sin^{2}{(\omega_{k,1}t)}\right] +\sin^{4}\theta \left[1 - \frac{m_{2}^{2}}{\omega_{k,2}^{2}}\sin^{2}{(\omega_{k,2}t)}\right] \notag \\
     && +  \frac{\sin^{2}{2\theta}}{2}  \bigg\{\cos{\omega_{k,1}t}\cos{\omega_{k,2}t} + \cos{2\sigma_{\bm k,1}}\cos{2\sigma_{\bm k,2}} \sin{\omega_{k,1}t}\sin{\omega_{k,2}t}\bigg\} \notag
\end{eqnarray}
or equivalently,
\begin{eqnarray}
     \braket{Q_{e}^{L}(t)} &=& 
    \cos^{4}\theta \left[1 - \frac{m_{1}^{2}}{\omega_{k,1}^{2}}\sin^{2}{(\omega_{k,1}t)}\right] +\sin^{4}\theta \left[1 - \frac{m_{2}^{2}}{\omega_{k,2}^{2}}\sin^{2}{(\omega_{k,2}t)}\right] \notag \\
     && +  \frac{\sin^{2}{2\theta}}{2}  \bigg\{\cos{\omega_{k,1}t}\cos{\omega_{k,2}t} + \frac{k^{2}}{\omega_{k,1}\omega_{k,2}} \sin{\omega_{k,1}t}\sin{\omega_{k,2}t}\bigg\} 
     \label{prob_survival}
\end{eqnarray}

In order to demonstrate that the expression for the average value of the chiral-flavor charge in Eq.~\eqref{prob_survival} matches the formula derived in~\cite{Bittencourt2021} for the survival probability of a neutrino with well-defined chirality (left) and flavor (electron), we add and subtract the term
\begin{eqnarray}
    \frac{m_{1}}{\omega_{k,1}}\frac{m_{2}}{\omega_{k,2}}\sin{\omega_{k,1}t}\sin{\omega_{k,2}t}
    + \frac{\sin^{2}{2\theta}}{2} \sin{\omega_{k,1}t}\sin{\omega_{k,2}t}\,,
\end{eqnarray}
 obtaining
\begin{eqnarray}
\frac{\sin^{2}{2\theta}}{2} \sin{\omega_{k,1}t}\sin{\omega_{k,2}t}
\end{eqnarray}
it is
\begin{eqnarray}
    \braket{Q_{e}^{L}(t)} &=& 
    \cos^{4}\theta + \sin^{4}\theta - \left[ \frac{m_{1}}{\omega_{k,1}}\cos^{2}{\theta} \sin{(\omega_{k,1}t)} + \frac{m_{2}}{\omega_{k,2}}\sin^{2}{\theta} \sin{(\omega_{k,2}t)}\right]^{2} \notag \\
     && +  \frac{\sin^{2}{2\theta}}{2}  \bigg\{  \left(\frac{k^{2}+m_{1}m_{2}}{\omega_{k,1}\omega_{k,2}} -1 \right)\sin{\omega_{k,1}t}\sin{\omega_{k,2}t}\bigg\} \notag \notag + \frac{\sin^{2}{2\theta}}{2} \left[ \cos{\omega_{k,1}t}\cos{\omega_{k,2}t}+\sin{\omega_{k,1}t}\sin{\omega_{k,2}t} \right].
\end{eqnarray}
By considering that
\begin{eqnarray}
\cos{\omega_{k,1}t}\cos{\omega_{k,2}t}+\sin{\omega_{k,1}t}\sin{\omega_{k,2}t} = \cos{[(\omega_{k,2}-\omega_{k,1})t]} = 1 - 2  \sin^{2}{\left[\frac{(\omega_{k,2}-\omega_{k,1})t}{2} \right]}
\end{eqnarray}
we have
\begin{equation}
    P_{e \to e}(t) = P_S^{e \to e}(t) + A_e(t) + B_e(t),
\end{equation}
where $ P_S^{e \to e}(t) $ is the standard flavor oscillation formula:
\begin{equation}
    P_S^{e \to e}(t) = 1 - \sin^2(2\theta) \sin^2 \left( \frac{\omega_{k,2}- \omega_{k,1}}{2} t \right),
\end{equation}
and
\begin{equation}
    A_e(t) = - \left( \frac{m_1}{\omega_{k,1}} \cos^2(\theta) \sin(\omega_{k,1} t) + \frac{m_2}{\omega_{k,2}} \sin^2(\theta) \sin(\omega_{k,2} t) \right)^2,
\end{equation}
\begin{equation}
    B_e(t) = \frac{1}{2} \sin^2(2\theta) \sin(\omega_{k,1} t) \sin(\omega_{k,2} t) 
    \left( \frac{k^2 + m_1 m_2}{\omega_{k,1} \omega_{k,2}} - 1 \right).
\end{equation}

\end{document}